\pdfoutput=1
\documentclass[preprint,groupedaddress,nofootinbib]{revtex4-1}

\usepackage{amsfonts}
\usepackage{amsmath}
\usepackage{graphicx}
\usepackage{hyperref}
\usepackage{epsfig}
\usepackage{verbatim}   % useful for program listings
\usepackage[subfigure]{graphfig}
\usepackage{epstopdf}
\usepackage{dcolumn}% Align table columns on decimal point
\usepackage{color}
\usepackage{tabularx}
\usepackage{braket}
\usepackage{float}
\usepackage{tikz}
\usepackage[percent]{overpic}
\usepackage{subfigure}
\usepackage{mwe}
\usepackage{soul}
\usepackage{ulem}

\usepackage[scale=0.8,bmargin=3cm,footnotesep=2cm]{geometry}
\usepackage[flushmargin]{footmisc}

\hypersetup{colorlinks, linkcolor = [rgb]{0,0.0,0.75}, citecolor = [rgb]{0,0.0,0.75}, urlcolor = [rgb]{0,0.0,0.75}}

\usepackage[utf8]{inputenc}
\usepackage{lineno}
\def\be{\begin{equation}}
\def\ee{\end{equation}}
\def\bea{\begin{eqnarray}}
\def\eea{\end{eqnarray}}
\newcommand{\Tr}{\mathrm {Tr}}

\newcommand{\nn}{\nonumber}
\newcommand{\la}{\label}

\definecolor{green}{rgb}{0,.5,0}

\begin{document}

\preprint{MSUHEP-17-005}

\title{Sea Quarks Contribution to the Nucleon Magnetic Moment and Charge Radius at the Physical Point}

\author{{Raza Sabbir Sufian$^{1}$, Yi-Bo Yang$^{1,2}$, Jian Liang$^{1}$, Terrence Draper$^{1}$, Keh-Fei Liu$^{1}$ }
\vspace*{-0.5cm}
\begin{center}
\large{
\vspace*{0.2cm}
\includegraphics[scale=0.15]{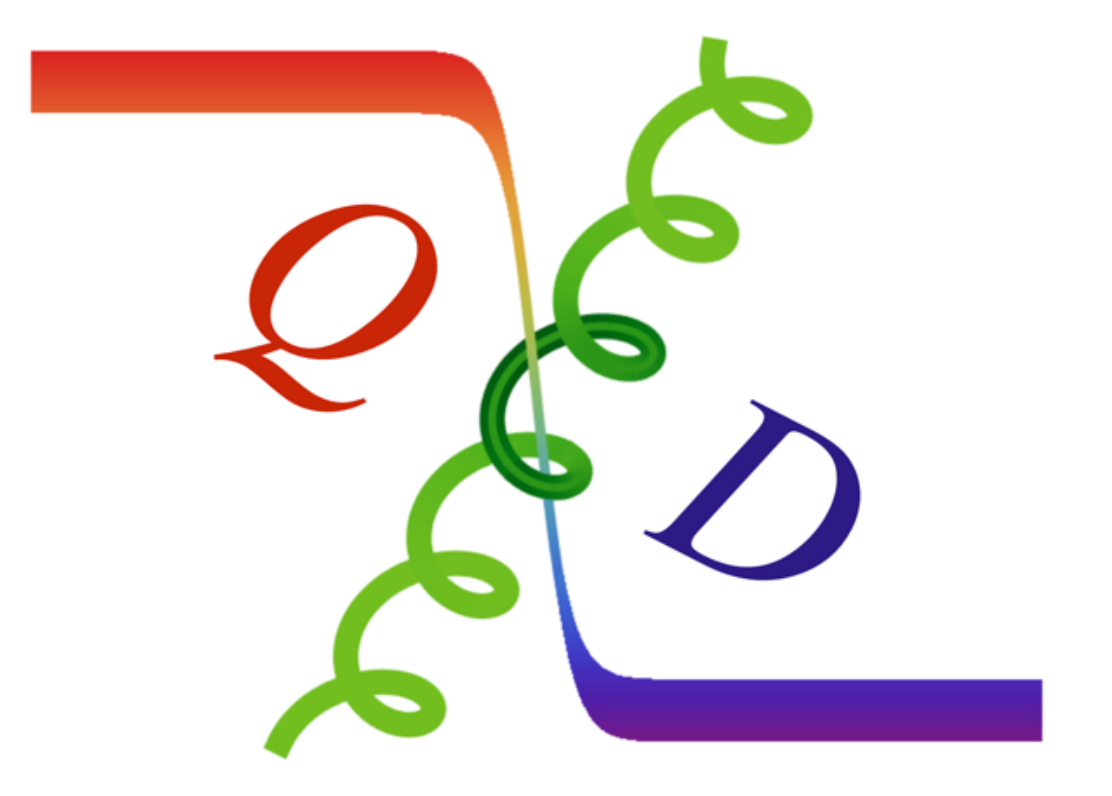}\\
\vspace*{-0.5cm}
($\chi$QCD Collaboration)
}
\end{center}
}
\vspace*{-0.5cm}
\affiliation{
$^{1}$\mbox{Department of Physics and Astronomy, University of Kentucky, Lexington, Kentucky 40508, USA}\\
$^{2}$\mbox{Department of Physics and Astronomy, Michigan State University, East Lansing, Michigan 48824, USA}
}

%\linenumbers

\begin{abstract}
We report a comprehensive analysis of the light and strange disconnected-sea quarks contribution to the nucleon magnetic moment, charge radius, and the electric and magnetic form factors. The lattice QCD calculation includes ensembles across several lattice volumes and lattice spacings with one of the ensembles at the physical pion mass. We adopt a model-independent extrapolation of the nucleon magnetic moment and the charge radius. We have performed a simultaneous chiral, infinite volume, and continuum extrapolation in a global fit to calculate results in the continuum limit. We find that the combined light and strange disconnected-sea quarks contribution to the nucleon magnetic moment is \mbox{$\mu_M\,(\text{DI})=-0.022(11)(09)\,\mu_N$} and to the nucleon mean square charge radius is \mbox{$\langle r^2\rangle_E\,\text{(DI)}=-0.019(05)(05)$ fm$^2$} which is about $1/3$ of the difference between the $\langle r_p^2\rangle_E$ of electron-proton scattering and that of a muonic atom and so cannot be ignored in obtaining the proton charge radius in the lattice QCD calculation. The most important outcome of this lattice QCD calculation is that while the combined light-sea and strange quarks contribution to the nucleon magnetic moment is small at about $1\%$, a negative $2.5(9)\%$ contribution to the proton mean square charge radius and a relatively larger positive $16.3(6.1)\%$ contribution to the neutron mean square charge radius come from the sea quarks in the nucleon. For the first time, by performing global fits, we also give predictions of the light and strange disconnected-sea quarks contributions to the nucleon electric and magnetic form factors at the physical point and in the continuum and infinite volume limits in the momentum transfer range of $0\leq Q^2\leq 0.5$ GeV$^2$.  
 
\end{abstract}

\maketitle

\section{Introduction} \la{intro}
Nucleon electromagnetic form factors of a hadron are of substantial interest because they are related to the dynamical content of the electric and magnetic currents distribution inside the hadron and characterize the internal structure of a nonpointlike particle. The quest for a detailed quantitative understanding of the nucleon electromagnetic form factors is an active field of the experimental nuclear physics, lattice QCD simulations, and other model calculations. However, some unsolved questions still remain regarding the nucleon electromagnetic form factors and their properties at low momentum transfer $(Q^2)$. Detailed reviews of various experimental results and model calculations can be found in~\cite{Pacetti:2015iqa,Punjabi:2015bba} and the references therein. The most recent surprising discrepancy of the proton charge radius measured from the Lamb shift in muonic hydrogen~\cite{Pohl2010,Antognini2013} differs by more than $5\sigma$ from the radius extracted with 1$\%$ precision using the electron-proton scattering measurements and hydrogen spectroscopy. While the current Committee on Data for Science and Technology (CODATA) value of proton charge radius is $r^p_E=0.8751(61)$ fm~\cite{PDG}, the most recent muonic hydrogen Lamb shift experiment measures $r^p_E=0.84087(39)$ fm~\cite{Pohl1:2016xoo} which is 4$\%$ smaller than, and differs by $7\sigma$ from the CODATA value. Other than the possibility that one of the proton charge radius extractions is wrong or involves considerable systematic uncertainties, the consequence of the ``proton charge radius puzzle" can have serious impacts such as a new physics signature, anomalous QCD corrections, a $5\sigma$ adjustment of the Rydberg constant (in the absence of new physics explanations) which is measured with an accuracy of about 5 parts per trillion, and/or a revision of sources of systematic uncertainties in the measurements of neutrino-nucleus scattering observables. Recent results and reviews of the proton charge radius puzzle can be found in Refs.~\cite{Lee:2015jqa,Carlson:2015jba,Hill:2017wzi}.

A complete first-principles lattice QCD calculation of the nucleon magnetic moment and charge radius including both the valence and the connected-sea quarks, called connected insertion (CI), and the disconnected-sea quarks contribution, called disconnected insertion (DI), is of immense importance and is not yet present in the literature. By disconnected insertions, we mean the nucleon matrix elements involving self-contracted quark graphs (loops), which are correlated with the valence quarks in the nucleon propagator by the fluctuating background gauge fields. It has also been found in various experiments that nonvalence components in the nucleon hold surprisingly large effects in describing its properties. One desires to perform a simulation at the physical pion mass and consider large volumes and small lattice spacings and overall to obtain a very good signal-to-noise ratio to compare the lattice results with the experimental value -- which is a highly ambitious goal of the lattice QCD community with current numerical resources. In two previous lattice QCD calculations~\cite{Abdel-Rehim:2013wlz,Green:2015wqa} the authors have calculated the light disconnected-sea quarks contribution to the nucleon electromagnetic form factors. In Ref.~\cite{Abdel-Rehim:2013wlz}, the simulation has been done with quark mass equivalent to pion mass 370 MeV and the authors obtained a light disconnected-sea quarks contribution to the nucleon electromagnetic form factor (EMFF) consistent with zero within uncertainties. In Ref.~\cite{Green:2015wqa}, the light disconnected-sea quarks contribution to the nucleon EMFF was obtained to be nonzero in the momentum transfer range of $0\leq Q^2\lesssim 1.2$ GeV$^2$ with the simulation performed at a quark mass equivalent to pion mass 317 MeV.

The light disconnected-sea quarks contribution to the nucleon EMFF has not been considered in most of the lattice calculations because of the following reasons: 1) the current status of the lattice QCD simulations with disconnected quark loops are numerically intensive and in general very noisy, especially near the physical pion mass, and 2) most of the previous lattice QCD calculations were performed under the assumption that DI light quarks contribute a negligible amount to the nucleon magnetic moment and charge radius. Therefore, most of the earlier simulations aimed to calculate only the isovector nucleon quantities and simulations were performed at relatively heavier pion masses~\cite{Yamazaki:2009zq,Capitani:2015sba,Lin:2010fv,Alexandrou:2011db,Alexandrou:2016rbj,Bhattacharya:2013ehc,Syritsyn:2009mx,Djukanovic:2015hnh}. Since gauge configurations with simulations directly at the physical pion mass are now becoming available, some collaborations are pursuing lattice QCD calculations near or at the physical pion mass. Nonetheless, simulations near the physical pion mass exhibit increased sensitivity to statistical fluctuations and one requires a large number of measurements to obtain a good signal-to-noise ratio and to control the undesired excited-states contaminations. Thus a majority of the recent calculations near the physical pion mass still concentrates on the CI calculations only~\cite{Syritsyn:2015nla,Abdel-Rehim:2015jna,Alexandrou:2017msl,Green:2014xba}.  

By performing a first-principles calculation, we find that the total contribution of the light (up and down) and strange disconnected-sea quarks to the nucleon mean square charge radius is negative and significant. Combining the result of the strange quark magnetic moment and charge radius calculated in our previous work~\cite{Sufian:2016pex} with the DI light-quarks contribution, we obtain the total contribution to the nucleon magnetic moment and mean square charge radius from the disconnected-sea quarks. Our overall DI calculation uncertainty is large compared to the precision of the experimental measurement of the proton charge radius and one also needs to perform a CI calculation at the physical point with high precision to draw any conclusion as to whether the DI contribution has a significant impact on the understanding of the $4\%$ discrepancy of the proton charge radius puzzle from the lattice QCD viewpoint. Nonetheless, the present work gives the first calculation of the light and strange disconnected-sea quarks contribution to the nucleon EMFF at the physical point and provides important information about the sign of the disconnected-sea quarks contribution to the nucleon EMFF. While almost all lattice QCD connected-insertion calculations concentrate on extractions of the proton charge radius, the neutron Sachs electric form factor $G^n_E(Q^2)$ calculation is challenging due to the poor signal-to-noise ratio, as shown in Ref.~\cite{Tang:2003jh}. A recent lattice QCD calculation~\cite{Alexandrou:2016hiy} performed at the physical pion mass also shows that obtaining a precise prediction of $G^n_E(Q^2)$ and neutron charge radius close to the experimental value is indeed a challenging problem. In this work, we have investigated the importance of the DI contribution to the neutron electric form factor calculation and a clear message is to be taken that one must include the DI contribution to the neutron charge radius to shift the lattice estimates toward the experimental value. It also gives a non-negligible contribution to the proton charge radius.

 This paper is organized as follows: an overview of the simulation details and statistics used in this work is provided in Sec.~\ref{simulation}. In Sec.~\ref{CF}, we provide examples of a hybrid two-states fit to compute matrix elements from the ratio of nucleon three-point to two-point correlation functions. We implement a model-independent extrapolation of a nucleon magnetic moment and a mean square charge radius from the EMFFs in the momentum transfer range of $0.051\lesssim Q^2\lesssim 1.31$ GeV$^2$ and show examples in Sec.~\ref{zexpEMFF}. In Sec.~\ref{results}, finite lattice spacing and finite volume corrections are included in a global fit with 24 valence quark masses on four different lattice ensembles with different lattice spacings, different volumes, and four sea quark masses including one at the physical point. From the fit coefficients of the model-independent $z$-expansion, we perform global fits to get estimates of the light and strange disconnected-sea quarks contributions to the nucleon electromagnetic form factors at the physical point. Finally, we present a conclusion to our lattice QCD analysis in Sec.~\ref{conclude}.

\section{Simulation Details} \la{simulation} 
Our calculation comprises numerical computation with a valence overlap fermion on four ensembles of $(2+1)$ flavor RBC/UKQCD domain-wall fermion (DWF) gauge configurations. We use 24 valence quark masses in total for the 24I, 32I, 32ID, and 48I ensembles corresponding to pion masses in the range $m_{\pi}\in$(0.135, 0.403) GeV to explore the quark-mass dependence of the DI EMFFs. Details of these ensembles can be found in Table~\ref{table:r0}. 
\begin{table}[htbp]
\begin{center}
%\vspace{10 mm}
\begin{tabular}{|c|c|c|c|c|}
\hline
Ensemble & $L^3\times T$  &$\mathnormal{a}$ [fm] &   {$m_{\pi}$} [GeV]  & $N_\text{config} $ \\
\hline
24I~\cite{Aoki:2010dy} & $24^3\times 64$& 0.1105(3) &0.330  & 203    \\
\hline
32I~\cite{Aoki:2010dy} &$32^3\times 64$& 0.0828(3)    &0.300 & 309 \\
\hline
32ID~\cite{Blum:2014tka} &$32^3\times 64$& 0.1431(7)& 0.171 & 200\\
\hline
48I~\cite{Blum:2014tka} &$48^3\times 96$& 0.1141(2)   &0.139 & 81 \\
\hline
\end{tabular}
\end{center}
\caption{\label{table:r0} The parameters for the DWF ensembles: spatial and temporal size, lattice spacing, the pion mass corresponding to the degenerate light-sea quark mass and the numbers of configurations used in this work.}
\end{table}
For the 24I and 48I lattices, we use 12\,-\,12\,-\,12\,-\,32 (16\,-\,16\,-\,16\,-\,32 for 32I and 32ID) random $Z_3$-noise grid sources with Gaussian smearing. Here, the first three numbers in the notations such as 12\,-\,12\,-\,12\,-\,32 denote the intervals of the grid in the 3-spatial directions and the last number is the interval between time slices. Therefore, on the 24I ensemble, the number of points in the grid has the pattern of 2\,-\,2\,-\,2\,-\,2. We place two nucleon sources on the time slices $t=0$ and $t=32$ and perform the inversion simultaneously. In addition, we repeat the inversion multiple times, shifting these nucleon sources in every two-time slice and therefore have 32 nucleon sources with 8 sets of stochastic noises for each of the 16 inversions on different time slices to tie the three quarks together for each smeared source. Therefore, the number of measurements for one configuration on the 24I ensemble is  $N_\text{grids}\times N_\text{sources}=(2^3\times 2)\times 32$. Finally, for the 203 configurations of the 24I ensemble, we have in total $(2^3\times 2)\times  32 \times 203 = 103936$ measurements from $1\times 32 \times 203 =6496$ inversions. Similarly, the number of measurements and the number of inversions on the 32I and 32ID ensembles per configuration are the same as those on the 24I ensemble. With the grid pattern of 4\,-\,4\,-\,4\,-\,3 on the 48I ensemble, the nucleon sources placed at time slices $t=0,\,32,64,$ and these sources shifted in every three-time slices, the number of measurements with 81 gauge configurations is $4^3\times 3\times 32 \times 81 = 497664$ and the number of inversions is $1\times 32 \times 81=2592$. A more detailed explanation of the grid source and the smearing can be found in Ref.~\cite{Liang:2016fgy}. We apply eigenmode deflation during the inversion of the quark matrix and utilize the low-mode substitution technique developed in Refs.~\cite{Li:2010pw,Gong:2013vja} to construct the nucleon propagator. The low-frequency part of the hadron correlators constructed using low mode substitution makes the use of a grid source feasible; otherwise no extra statistics can be gained for the nucleon~\cite{Li:2010pw}. As for the quark loops, the low-mode part is exactly calculated with the low eigenmodes of the overlap operator which is called low-mode average, and the high-mode part is estimated by 8 sets of 4\,-\,4\,-\,4\,-\,2 $Z_4$ noise grids with even-odd dilution as well as additional time dilution~\cite{Gong:2013vja,Yang:2015uis}. The noise-estimated high-mode part requires the calculation of two noise propagators for the even-odd spatial dilution and another two noise propagators for the time dilution, repeating these inversions for 8 grids. Therefore, with 8 different sets of $Z_4$ noise grids, we have to perform $2\times 2\times 8 =32$ inversions. With these techniques implemented, our statistics are from $\sim100\text{k}$ to $\sim500\text{k}$ measurements on the 24I to 48I ensembles. 

We define the nucleon two-point (2pt) and three-point (3pt) correlation functions as
 \bea
 \Pi^{2pt}(\vec{p}\,',\!t_2;\!t_0)\!&\equiv&\!\sum_{\vec{x}}\!e^{-i\vec{p}\,'\cdot\vec{x}} \!\Bra{0}\!T[\chi(\vec{x},\!t_2) \!\sum_{x_i\!\in G}e^{\vec{p}\,'\cdot \vec{x}_i}\,\bar{\chi}_{S}(x_i,\!t_0)]\! \Ket{0}, \nn \\
 \Pi^{3pt}_{V_\mu} (\vec{p}\,'\!, t_2;\! \vec{q},\! t_1;\!t_0) \!&\equiv&\! \sum_{\vec{x}_2, \vec{x}_1}\! e^{-i\vec{p}\,'\cdot\vec{x}_2+i\vec{q}\cdot\vec{x}_1}\! \Bra{0}T[\chi(\vec{x}_2,\!t_2)  V_\mu(\vec{x}_1,\!t_1)\!\sum_{x_i\in G}\!\bar{\chi}_{S}(x_i,\!t_0)] \!\Ket{0} ,
 \eea
where $t_0$ and $t_2$ are the source and sink temporal positions, respectively, $\vec{p}\,'$ is the sink momentum, respectively, and $t_1$ is the time at which the bilinear operator $V_\mu(x)=\bar{q}(x)\gamma_\mu q(x)$ is inserted with $q$ a light (up or down) or strange quark. $x_i$ are points on the spatial grid $G$ (as described above), $\chi$ is the usual nucleon point interpolation field and $\bar{\chi}_{S}$ is the nucleon interpolation field with smeared stochastic grid source ($Z_3$-noise source), and the three-momentum transfer is $\vec{q}=\vec{p}\,'-\vec{p}$ with $\vec{p}$ the source momentum. For the point sink and smeared source with $t_0=0,$ $\vec{p}=\vec{0},$ and $\vec{q}=\vec{p}\,'$, the Sachs FFs can be obtained by the ratio of a combination of 3pt and 2pt correlations with appropriate kinematic factors,
\bea \la{RatioEq}
R_\mu(\vec{q},t_2, t_1) \equiv \frac{\Tr [\Gamma_m \Pi^{3pt}_{V_\mu} (\vec{q},t_2, \!t_1 ) ]} {\Tr [\Gamma_e \Pi^{2pt} (\vec{0},t_2)]} e^{(E_q-m)\cdot(t_2-t_1)} \frac{2E_q}{E_q+m}. 
\eea
Here, $E_q=\sqrt{m_N^2+\vec{q}\,^2}$ and $m_N$ is the nucleon mass. The choice of the projection operator for the magnetic form factor is $\Gamma_m \!=\!\Gamma_k\! =\! -i(1+\gamma_4)\gamma_k\gamma_5/2$ with $k\!=\!1,2,3$ and that for the electric form factor is $\Gamma_e\!=\!(1+\gamma_4)/2$. 

Notice that we use smeared grid source and point sink. We have performed numerical check on the 32ID ensemble to examine the signal-to-noise ratio of the smeared-smeared nucleon 3pt/2pt correlation function ratio to that of the smeared-point 3pt/2pt correlation function ratio. For this particular ensemble, at the unitary point (sea quark mass corresponding to $m_\pi=171\,\text{MeV}$), we find that the smeared-source and smeared-sink 3pt/2pt correlation function ratio is about 2\%\,-\,6.5\% noisier than the smeared-source and point-sink 3pt/2pt correlation function ratio in the time window where we perform two-states fit to obtain nucleon matrix elements. A careful check also shows that the smeared-smeared 2pt correlation function is only about 1\%\,-\,2.5\% noisier than the smeared-point 2pt correlation function in the time window where we perform fit to obtain nucleon effective mass, while the central value of the nucleon effective mass remains almost the same for both cases. Since the statistical uncertainty of the nucleon matrix elements near the unitary point on the 32ID ensemble is about $50\%$, therefore the final result would not be significantly different if we use the smeared-point or smeared-smeared 2pt correlation function in our calculation. Therefore, we have used the smeared-point two-point function for the numerical analysis in this work. Also, without much additional computational cost, we cannot implement the standard square-root technique to calculate the nucleon 3pt/2pt ratio. We use the smeared source for the three-point function which would invoke a smeared-smeared two-point function in the square-root formula. Since we use the smeared-source-point-sink three-point function, the factor $Z_p (q)/Z_p(0)$, where $Z_p(q)$ is the interpolation-field overlap factor for a point source with the nucleon momentum $q$, is not exactly canceled in the ratio defined in Eq.~\ref{RatioEq}. In the continuum limit, this extra factor is unity and, on the lattice, it will have a $q^2 a^2$ error which can be absorbed in the zero-momentum extrapolation of $G_M$ and charge radius and the subsequent continuum extrapolations. We have numerically checked on about 100 configurations on the 32I (smallest lattice spacing) and 32ID (largest lattice spacing) ensembles that the interpolation field overlap factors indeed do not cancel for nonzero momentum but have a small effect on the matrix element (typically 5\%\,-\,6\% for the largest momentum and the lightest pion mass). Upon performing the $z$-expansion~\cite{Hill:2010yb,Epstein:2014zua} to obtain the magnetic moment at $Q^2=0$, the effect on the final result is even smaller, about 1\%\,-\,2\%. The charge radius calculated with such correction has a change of about $2\%$ on the 32I ensemble and $1\%$ on the 32ID ensemble lattice results. Since our statistical uncertainty is about $25\%$ in the global fit for the magnetic moment and the charge radius with an additional $10\%$ (for magnetic moment) and $20\%$ (for charge radius) systematic uncertainties from the $z$-expansion results will be included in the final result of the global fits, this small effect of interpolation-field overlap factors does not affect our calculation in a significant way. For the 32ID and 48I ensembles, the $Q^2$ are much smaller than those of 24I and 32I ensembles and the overlap ratio itself is at the 1\%\,-\,2\% level. We thus ignore it in order to reduce additional computational costs.

In the limits $(t_2 - t_1) \gg 1/\Delta m$ and $t_1 \gg 1/\Delta m$, one can obtain the Sachs magnetic and electric FFs by an appropriate choice of projection operators and current directions $\mu$,
\bea
R_{\mu = i} (\Gamma_k) \xrightarrow{(t_2 - t_1) \gg 1/\Delta m, t_1 \gg 1/\Delta m} &&  \frac{\epsilon_{ijk}q_j}{E_q+m_N} G_M(Q^2),\nn \\
R_{\mu = 4} (\Gamma_e) \xrightarrow{(t_2 - t_1) \gg 1/\Delta m, t_1 \gg 1/\Delta m}&& G_E(Q^2),
\eea
with ${i,j,k}\neq4$ and $\Delta m$ the mass gap between the ground state and the first excited state. The Sachs magnetic and electric FFs in the spacelike region are related to the nucleon Dirac ($F_1$) and Pauli FF ($F_2$) through the relations
\bea
G_M(Q^2) &=&F_1(Q^2)+F_2(Q^2) \nn \\
G_E(Q^2) &=&F_1(Q^2)-\frac{Q^2}{4m_N^2}F_2(Q^2).
\eea

\section{Combined two-states fit} \la{CF}
In lattice QCD simulations, nucleon correlation functions suffer from an exponentially increasing noise-to-signal ratio which imposes a serious limitation on the source-sink separation $t_2$, especially when DI calculations are performed. In general, DI calculations are notoriously noisier compared to the CI calculations. It is also hard to extract the ground-state properties of the nucleon since the lowest excited state, the Roper resonance $N(1440)$ lies close to the nucleon mass. There can also be an additional excited-states contamination, for example from the $\pi N$-states. Therefore, ideally one requires a substantially large source-sink separation, approximately $t_2 =1.5$ fm, to extract nucleon ground-state matrix elements so as not to be much affected by the excited-states contamination. Though it is possible to go up to about $1.4$ fm source-sink separation in some of the CI calculations~\cite{Yamazaki:2009zq,Green:2014xba} only, at the present stage of numerical simulation it is quite challenging to go much beyond $t_2\approx$ 1\,fm and obtain a reasonable signal-to-noise ratio for the DI calculations. Therefore, to have an estimate of the nucleon ground-state matrix elements, we employ a hybrid joint two-state correlated fit by simultaneously fitting the standard 3pt/2pt ratio $R(t_2,t_1)$ and the widely used summed ratio $SR(t_2)$~\cite{Maiani:1987by} to calculate DI matrix elements. The $R(t_2, t_1)$ and $SR(t_2)$ fitting formulas for a given direction of current and momentum transfer can be written, respectively, as~\cite{Yang:2015zja}
\bea \la{CF-method1}
R(t_2,t_1) = C_0 + C_1 e^{-\Delta m(t_2-t_1)} +C_2 e^{-\Delta m t_1} +C_3 e^{-\Delta m t_2},
\eea
\bea \la{CF-method2}
SR(t_2) &\equiv&  \sum_{t_1 \geq t'}^{t_1\leq (t_2-t'')} R(t_2,t_1)\nn \\
&=&(t_2-t' - t^{''} +1)C_0 + C_1 \frac{e^{-\Delta m t''}-e^{-\Delta m (t_2-t' + 1)}}{1-e^{-\Delta m }} \nn \\
&+& C_2 \frac{e^{-\Delta m t'}\!-\!e^{-\Delta m (t_2-\!t'' \!+\! 1)}}{1-e^{-\Delta m }}+\! C_3(t_2-\!t' -\!t''+\!1)e^{-\Delta m t_2}. 
\eea
Here, $t'$ and $t^{''}$ are the number of time slices we drop at the source and sink sides, respectively, and we choose $t'=t''=1$. $C_i$ are the spectral weights involving the excited states, and $\Delta m$ is, in principle, the energy difference between the first excited state and the ground state. Basically, the two-states fit in Eq.~(\ref{CF-method1}) dominates in our combined fit method and, for heavier pion masses, the final result of the combined fit is almost identical to the standard 3pt/2pt ratio two-states fit. However, the combined fit becomes useful for getting a stable fit near the physical pion mass, and we gain a slight increase in the signal-to-noise ratio. We choose $t'$ and $t''=1$ by following the strategy of keeping as many points as possible for which $\chi^2$ is acceptable. We do not obtain any signal for the fit parameter $C_3$ based on the analysis of our lattice data points for light-sea quarks. Therefore, excluding this factor from the combined fit does not affect the final outcome of the fit. $\Delta m$ is effectively an average of the mass difference between the proton and the lowest few excited states and needs to be determined by the fit.

\begin{figure}[htbp] 
  \centering
  \subfigure[Two-states 3pt/2pt ratio fit]{\includegraphics[width=0.48\textwidth]{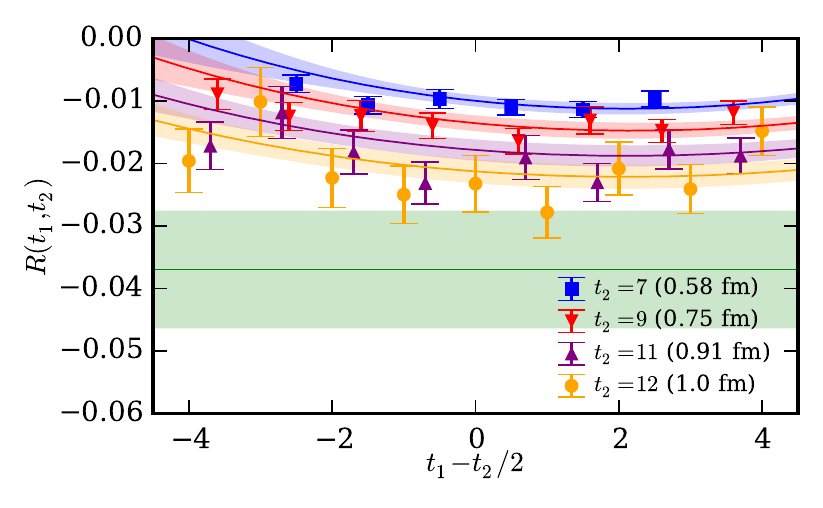}}
  \subfigure[Summed ratio fit]{\includegraphics[width=0.48\textwidth]{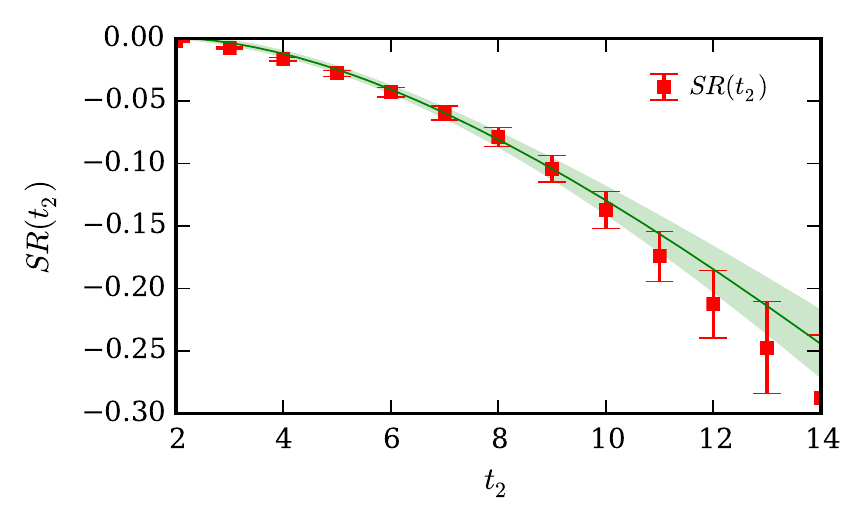}}
  \caption{\label{fig1} Combined correlated two-states fit of the 32I ensemble 3pt/2pt-ratio and summed ratio data. The transparent bands show the fit results based on the fit parameters listed in Table~\ref{table:r1}. The green bands show the final fit result of the light disconnected-sea quarks magnetic form factor $G^\text{light-sea}_M(Q^2)$ at $Q^2=0.218$ GeV$^2$. }
\end{figure}

\begin{figure}[htbp] 
  \centering
  \subfigure[Two-states 3pt/2pt ratio fit]{\includegraphics[width=0.48\textwidth]{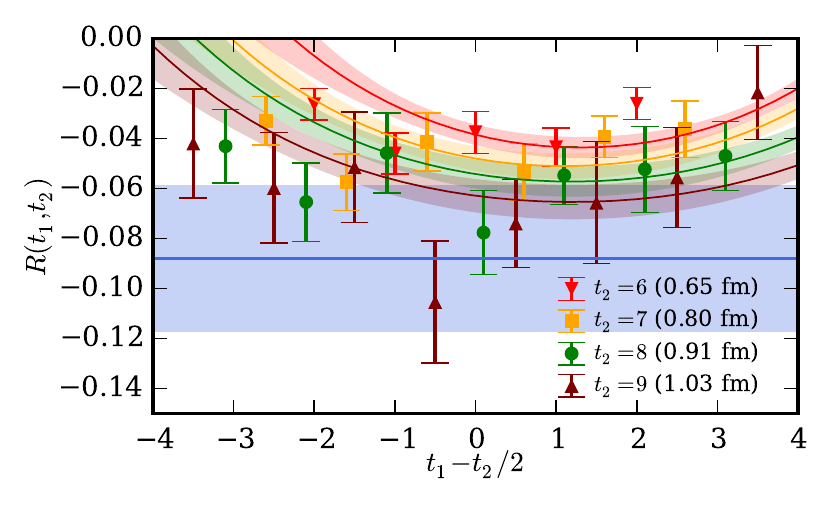}}
  %\hfill
  \subfigure[Summed ratio fit]{\includegraphics[width=0.48\textwidth]{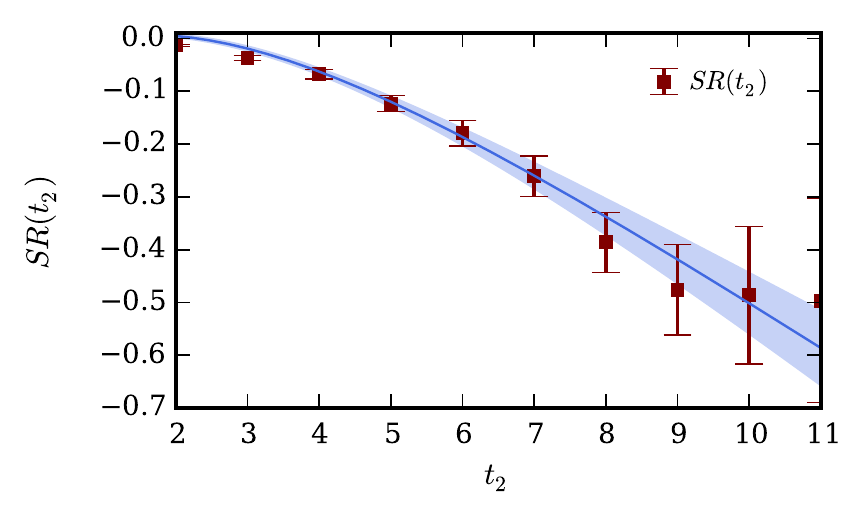}}
  \caption{\label{fig2} Combined correlated two-states fit of the 48I ensemble 3pt/2pt-ratio and summed ratio data. The transparent bands show the fit results based on the fit parameters listed in Table~\ref{table:r1}. The blue bands show the final fit result of the light disconnected-sea quarks magnetic form factor $G^\text{light-sea}_M(Q^2)$ at $Q^2=0.051$ GeV$^2$. }
\end{figure}

We illustrate two examples in Figs.~\ref{fig1} and \ref{fig2} to obtain magnetic form factors at given $Q^2$-values from the lattice data and present the fitting details in Table~\ref{table:r1}. The source-sink separation we use for the fitting of 32I ensemble data is $t_2\in{(6,13)}$ and $t_2\in{(5,10)}$ for the 48I ensemble data. As discussed earlier, as with almost all of the DI calculations, we are forced to constrain the $t_2$-window around $1.1$ fm due to the limitations of good signal-to-noise ratio. However, in principle, the two-states fit should compensate for this limitation to a certain degree. We perform a correlated combined fit of the ratio and summed ratio data. Likewise, all of the subsequent fits in the article are also correlated fits.

\begin{table}[htbp]
\begin{center}
%\vspace{10 mm}
\begin{tabular}{|c|c|c|c|c|c|c|c|c|}
\hline
Ensemble & $m_\pi$ [GeV] & $Q^2$ [GeV$^2$] & $C_0$ & $C_1$ & $C_2$ & $\Delta m$ [GeV]   & $\chi^2/\text{d.o.f.}$ \\
\hline
32I & 0.330 & 0.218 & $-0.036(09)$ & $0.018(06)$ & $0.025(06)$ & $0.350(121)$   & $1.26(5)$ \\
\hline
48I & 0.207 & 0.051 & $-0.088(29)$ & $0.062(18)$ & $0.072(23)$ & $0.637(250)$   & $1.04(7)$ \\
\hline
\end{tabular}
\end{center}
\caption{\label{table:r1} The parameters of correlated combined two-states fits to obtain the light disconnected-sea quarks magnetic form factor at given momentum transfers.}
\end{table}

From the combined fit Eqs.~(\ref{CF-method1}) and (\ref{CF-method2}), it is seen that when $\Delta m$ is large, the data points for different source-sink separation should have overlap amongst themselves or the separation between them should be small. A comparison between the fit values of $\Delta m$ in Table~\ref{table:r1} and Figs.~\ref{fig1} and \ref{fig2} agrees with this assessment. It is seen from Fig.~\ref{fig1} that a smaller value of $\Delta m$ is consistent with the well separated data points with different sink-source separations on the 32I ensemble. One can see from Fig.~\ref{fig2} and $\Delta m = 0.637(250)\, \text{GeV}$ from Table~\ref{table:r1} that a larger value of the energy gap is consistent with the overlapping data points at different $t_2$, and therefore, the final fit result is closer to the plateau region of the data points at source-sink separation $t_2=9$ of the 48I ensemble lattice data. However, a clear understanding of why the $\Delta m$ fit value is larger for the data with smaller pion mass on the 48I ensemble than that of the heavier pion mass on the 32I ensemble is lacking at this moment. As mentioned before, this $\Delta m$ actually gives an effective measure of the energy difference between the nucleon ground state and a sum of several excited states whose energies are above the ground state. Since $\Delta m$ reflects a weighted sum of the excited states, we speculate that when the quark mass is low enough, multiple $\pi$ N and $\pi\pi$ N states start to appear and it
could give a higher effective $\Delta m$. Moreover, like most of the present-day lattice DI calculations, we are also limited by our statistics to go beyond a source-sink separation of 1.5 fm to extract nucleon ground state matrix elements and obtain a clearer understanding of the excited-states contamination.

We perform similar combined correlated two-states fits to obtain the DI Sachs electric form factor and ensure that the fit window is as large as possible; in most cases the $\chi^2/\text{d.o.f}$ is in the vicinity of 0.9\,-\,1.1. We choose the largest possible fit window as long as goodness of the fit is ensured and one can obtain a reasonable signal-to-noise ratio in the fits.

\section{Extraction of the DI magnetic moment and charge radius}
\begin{figure}[htbp]
\begin{center}
\includegraphics[height=5.0cm,width=15.0cm]{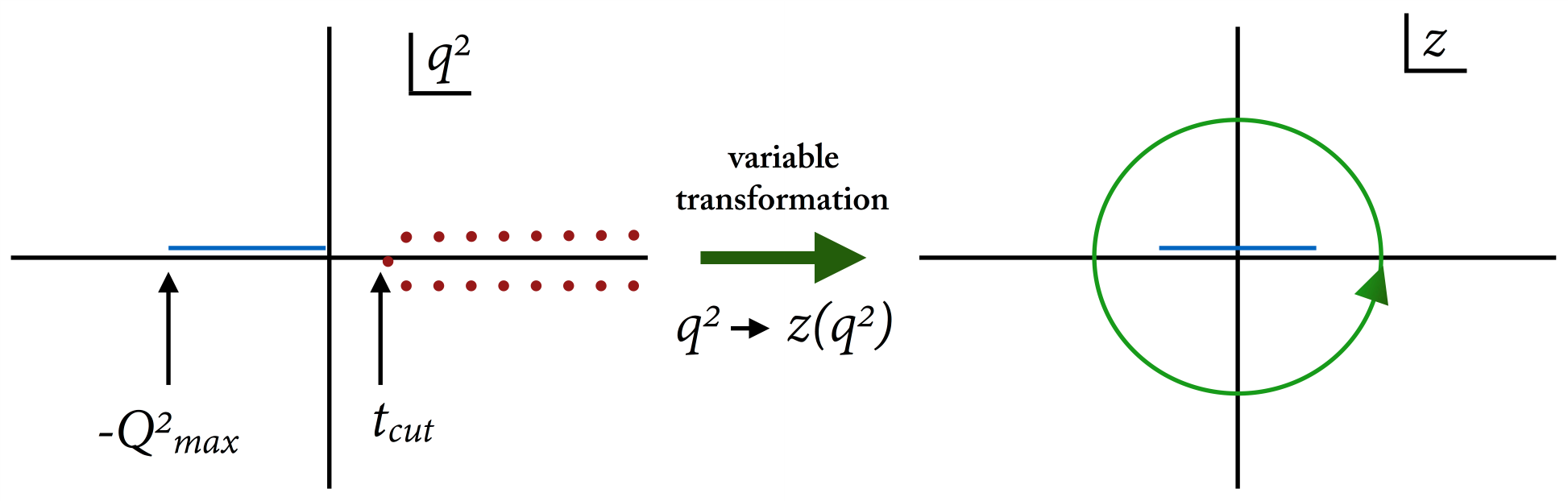}
\end{center}
\caption{\label{fig3}Model-independent $z$-expansion: Conformal mapping of the cut plane to the unit circle.}
\end{figure}

\la{zexpEMFF}
It has been a topic of long discussion about what type of form one should use to describe the $Q^2$-behavior of different form factors. A choice based on the phenomenological interpretation of various data is the dipole form~\cite{Hand:1963zz} which has been widely used. But a simple polynomial fit does not converge when there exist cuts in the timelike domain. For example, in the case of a photon to two pion transition, there exists a cut at $q^2=-Q^2=4m_\pi^2$ in the timelike domain as shown in Fig.~\ref{fig3}. Because of the existence of this pole $1/(q^2-4m_\pi^2)$, a polynomial expansion of the FF should not converge for any $Q^2> 4m_\pi^2$. The weight of this pole may be small, but one should not ignore its effect when fitting the FF data. To overcome this problem, a conformal mapping of variable $Q^2$ to another variable $z$ has been proposed in Refs.~\cite{Hill:2010yb,Epstein:2014zua}. The conformal mapping is performed in such a way that one is allowed to perform a polynomial expansion in $z$, such that the timelike momentum transfers (i.e. all poles of the FFs) map onto the unit circle $z=1$ and the spacelike momentum transfers map onto the real line $|z|<1$. For more details, see~\cite{Hill:2010yb,Epstein:2014zua}. 

Another reason we do not use the dipole fit in the calculation is because the $Q^2$ behaviors of the disconnected light and strange form factors are unknown and one would prefer not to be biased with a specific form of the extrapolation. (There exist also other phenomenological models for the $Q^2$-dependence of strange form factors, for example, in Ref.~\cite{Hemmert:1998pi}.) Therefore we adopt the model-independent $z$-fit. We take $t_\text{cut}=4m_\pi^2$ for fitting the light disconnected-sea quarks FF and $t_\text{cut}=4m_K^2$ for the strange quark FF. We have verified that a different choice of $t_\text{cut}$ such as $9m_\pi^2$ has less than a few percent effect on our extrapolations. 

\begin{figure}[htbp]
  \centering
  \subfigure[]{\includegraphics[width=0.48\textwidth]{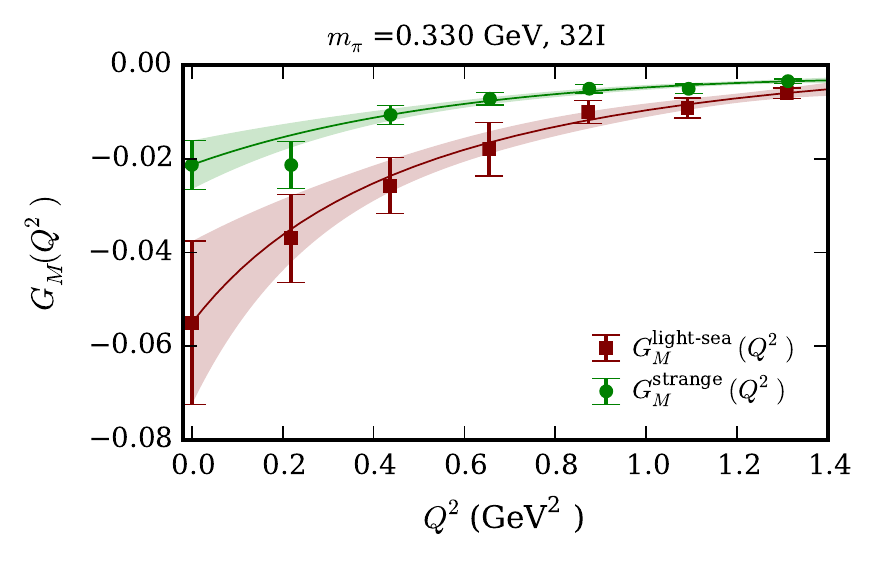}\la{4a}}
  \subfigure[]{\includegraphics[width=0.465\textwidth]{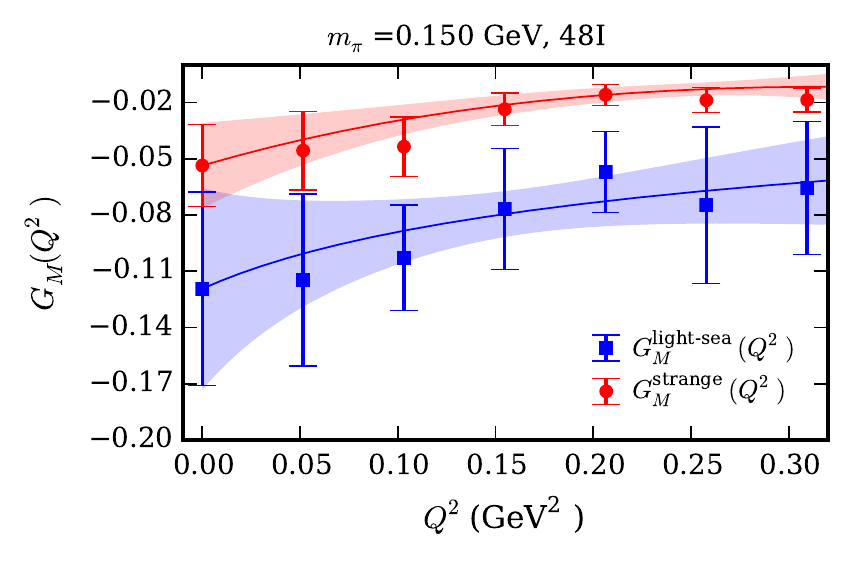}\la{4b}}
   \subfigure[]{\includegraphics[width=0.465\textwidth]{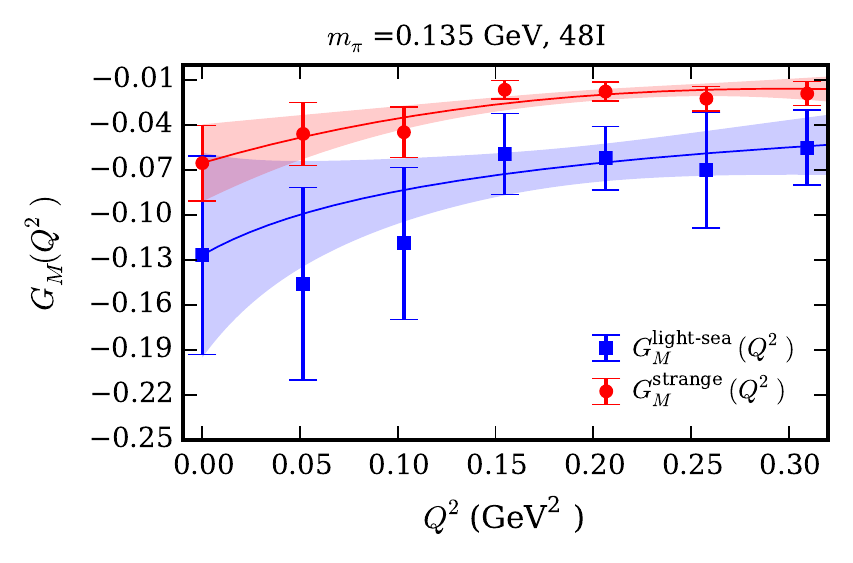}\la{4c}}
  \caption{\label{fig4}Light and strange disconnected-sea quarks magnetic moment $G^\text{light-sea,\,strange}_M(0)$ extrapolation for three different quark masses of the 32I (Fig.~(\ref{4a})) and 48I (Figs.~(\ref{4b}),~(\ref{4c})) ensembles using $z$-expansion from the lattice $G^\text{light-sea,\,strange}_M(Q^2)$. The $\chi^2/\text{d.o.f.}$ for the extrapolations are in the range of 0.520.88. Charge factors are not included in the form factors. Note the $Q^2$ ranges are different in the 32I and 48I cases.}
\end{figure}

In Fig.~\ref{fig4}, we show three examples of the extractions of light-sea-quarks magnetic moment at $Q^2=0$ from the FF data at different $Q^2$ using the $z$-fit,
\bea \la{zexpf}
G^{q,z-\text{exp}}_{E,M}(Q^2) &=& \sum^{k_\text{max}}_{k=0} a_k z^k,
\eea
where
\bea
z&=&\!\frac{\sqrt{t_{\text{cut}}+Q^2}-\sqrt{t_{\text{cut}}}}{\sqrt{t_{\text{cut}}+Q^2}+\sqrt{t_{\text{cut}}}}. \nn
\eea 
We see from Fig.~\ref{fig4} and also from our previous work~\cite{Sufian:2016pex} that the lattice data of the 48I ensemble is quite a bit noisier than the 24I and 32I ensemble data. Therefore we show in Figs.~(\ref{4b}) and (\ref{4c}) two examples of how we extract the light-sea and strange quarks contributions to the nucleon magnetic moment by performing simulation around the physical pion mass $m_\pi\in{(0.135,0.150)}$ GeV. 

As discussed in our previous work~\cite{Sufian:2016pex}, we keep the first 3-terms in the $z$-expansion formula~(\ref{zexpf}) and perform the $Q^2$-extrapolation. Unlike for the strange quark magnetic moment extraction in~\cite{Sufian:2016pex}, for the light disconnected-sea quarks magnetic moment, constraining $a_2$ with a prior width of $2\times |a_{2, \text{avg}}\vert$ does not have any effect since the uncertainties in the fit values of $a_2$ are already smaller than $2\times |a_{2, \text{avg}}\vert$ for almost all of the pion masses. Therefore we do not set any prior on $a_2$ for the extraction of the magnetic moments. However, for the extraction of the charge radii, we calculate the jackknife ensemble average $a_{2,\text{avg}}$ of the coefficient $a_2$ and then perform another fit by setting $a_2$ centered at $a_{2,\text{avg}}$ with a prior width equal to $2\times |a_{2, \text{avg}}\vert$. We find that the effect of setting this prior is almost insignificant for the 24I and 32I ensemble data, especially at heavier quark masses. However, the prior stabilizes the extrapolation of $G^q_E(Q^2)$ for pion masses around the physical point for the 48I ensemble. Since the $z$-expansion method guarantees that $a_k$ coefficients are bounded in size and that higher order $a_k$'s are suppressed by powers of $z^k$, we carefully check the effect of the $a_3$ coefficient in our fit formula and estimate this effect to calculate the systematic uncertainties in the $z$-expansion fits. We calculate the difference in the central values of $G^q_M(0)$ with and without the addition of the $a_3$ term for the lightest quark masses at the unitary point for each lattice ensemble. We find the addition of the $a_3$-term in the $z$-expansion after we constrain $a_2$ has the largest effect, as expected, for the quark mass equivalent to $m_\pi\sim140$ MeV of the 48I ensemble and obtain the difference in the central value of $G^\text{light-sea}_M(0)$ to be about 11\%. Therefore, we take a conservative approach and estimate a systematic error of 11\% of the final continuum value of $G^q_M(0)$ obtained from the global fit. 

\begin{figure}[htbp]
  \centering
  \subfigure[]{\includegraphics[width=0.48\textwidth]{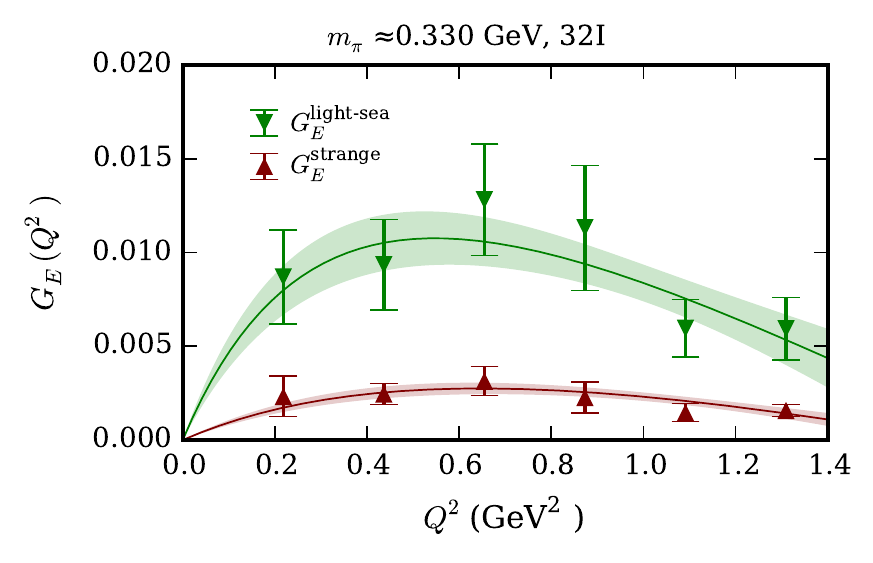}\la{5a}}
  \subfigure[]{\includegraphics[width=0.465\textwidth]{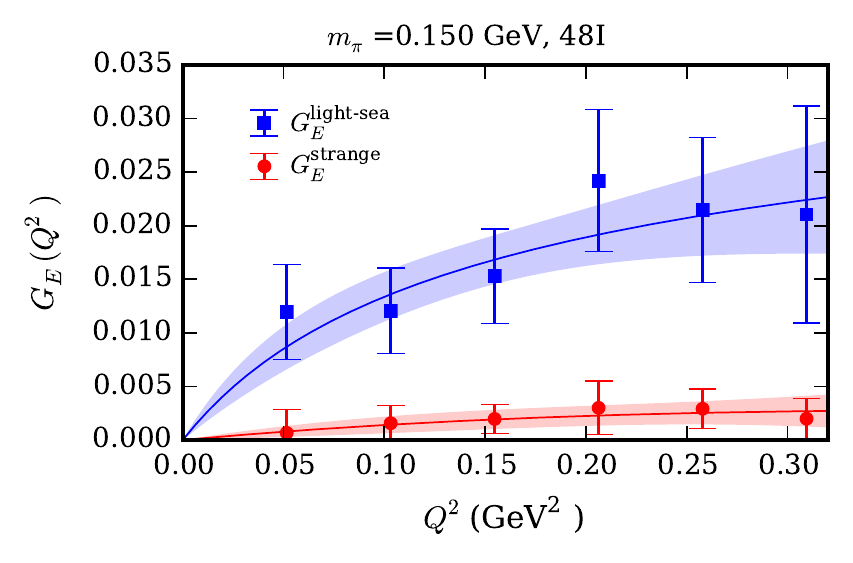}\la{5b}}
  \subfigure[]{\includegraphics[width=0.465\textwidth]{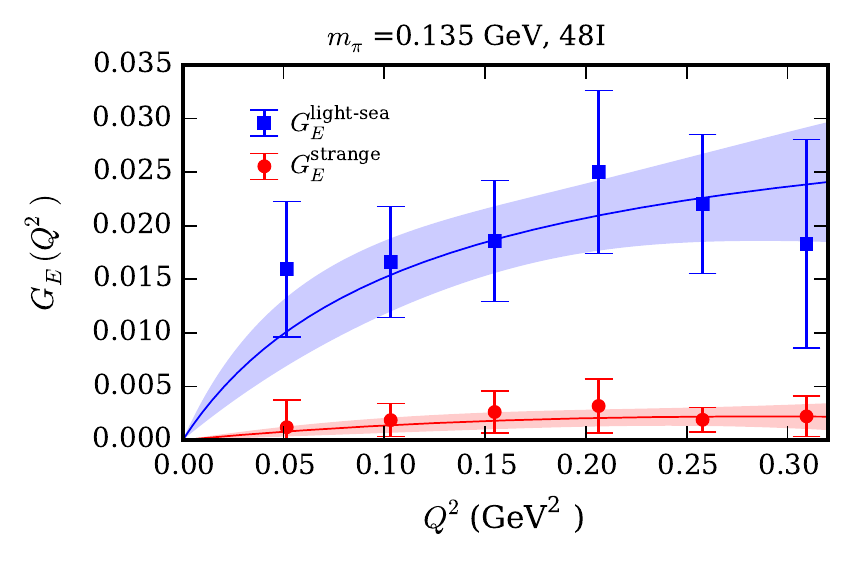}\la{5c}}
  \caption{\label{fig5}Light and strange disconnected-sea quarks contributions to the nucleon electric FF $G^\text{light-sea/strange}_E(Q^2)$ for two different quark masses of the 32I (Fig.~(\ref{5a})) and 48I (Figs.~(\ref{5b}),~(\ref{5c})) ensembles. The $\chi^2/\text{d.o.f.}$ for the two fits are in the range of 0.49\,-\,0.81. Charge factors are not included in the form factors. Note the $Q^2$ ranges are different in the 32I and 48I cases.}
\end{figure}

Similarly, one can extract the light and strange disconnected-sea quarks contributions to the nucleon charge radius by calculating the slope of $G^q_E(Q^2)$ near $Q^2=0$. We find that adding the $a_3$ term in the $z$-expansion has a larger effect on calculating the charge radius than in extracting the magnetic moment and such an effect of adding the $a_3$ term for the charge radius calculation is 12\%\,-\,20\%. Therefore a 20\% uncertainty has been added to the systematics in the global fit of charge radius as a part of our conservative assessment. One important observation from Fig.~\ref{fig5} is that although the data of light quark electric FF are not very precise, nevertheless the uncertainty band of the $z$-expansion is narrower compared to the magnetic FF extrapolation. The reason is due to charge conservation as the disconnected $G^q_E(Q^2)$ is constrained to be 0 at $Q^2=0$. Another important observation from Fig.~\ref{fig5} is that the light disconnected-sea quarks contribution to the $G^\text{light-sea}_E(Q^2)$ is almost 6\,-\,10 times larger than the strange quark contribution $G^{s}_E(Q^2)$. 
 
\section{Global fits of the disconnected insertions of nucleon properties} \la{results}

\begin{figure}[htbp]
\begin{center}
\includegraphics[height=6.0cm,width=9.0cm]{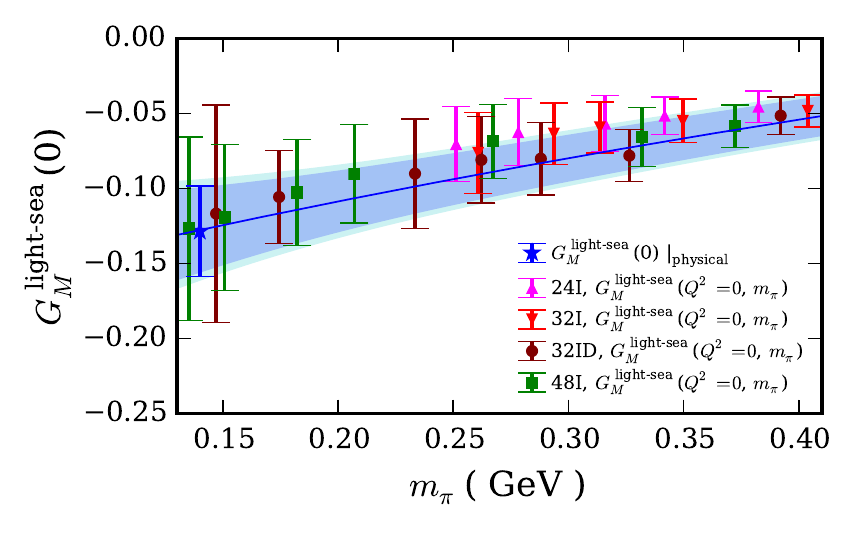}
\end{center}
\caption{\label{fig6}Light disconnected-sea quark magnetic moment at 24 quark masses on 24I, 32I, 48I, and 32ID ensembles as a function of the pion mass. The curved blue line shows the behavior in the infinite volume and continuum limit. The cyan band shows the combined statistical (blue band) and systematic uncertainties added in quadrature. The $\chi^2/\text{d.o.f.}$ of the fit is 0.67.}
\end{figure}

With the extrapolated results from the $z$-expansion in hand, we now have 24 data points for the magnetic moments and charge radii calculated from the slopes near $Q^2=0$ of the electric FFs. For the empirical global fit formula of the light-sea-quarks magnetic moment, we employ chiral extrapolation from Ref.~\cite{Jenkins:1992pi} and volume extrapolation from Ref.~\cite{Beane:2004tw}. One can add the $m_\pi^2\log(m_\pi^2)$ term~\cite{Jenkins:1992pi} in the chiral extrapolation of light disconnected-sea quarks magnetic moment, but we do not obtain any signal for this term by fitting the lattice data and the final value of the magnetic moment is independent of the addition of this term. Therefore we dropped this term from the chiral extrapolation of light-sea-quarks magnetic moment. Since the overlap fermion action is already $\mathcal{O}(a)$ improved, therefore, we apply an $\mathcal{O}(a^2)$ correction to the global fit formula
\bea\la{MFFglobal}
G^\text{light-sea}_M(Q^2&=&0,m_\pi,m_K,m_{\pi,vs},a,L) = A_0+A_1\,m_\pi  +A_2\, m_K \nn \\ 
&& +A_3\,a^2 +A_4\,m_\pi(1-\frac{2}{m_\pi\,L}) \,e^{-m_\pi L}
\eea
where $m_{\pi}\,(m_K)$ is the valence pion (kaon) mass, and $m_N$ is the nucleon mass. We show the extrapolation of the nucleon light disconnected-sea quarks magnetic moment in Fig.~\ref{fig6}. At the physical point and in the limit, {\it{i.e.}} $a\to 0$ and $L\to\infty$, we obtain
\bea\la{magmoment}
\left.G^\text{light-sea}_M(0)\right\vert_{\text{physical}} = - 0.129(30)(13)(18) \,\mu_N\, ,
\eea
where the magnetic moment is measured in the unit of nucleon magneton ($\mu_N$). The first uncertainty in the value of the magnetic moment in Eq.~(\ref{magmoment}) comes from the statistics, the second uncertainty comes from adding the higher order $a_3$-term in the $z$-expansion, and the third uncertainty comes from the variation of the central value in the global fit formula with the introduction of additional terms. The parameter values we obtain according to the global fit are $A_1=0.38(12),\, A_2=-0.40(16),\,A_3=0.30(39),\,A_4=-1.26(2.75)$. An attempt to add a partial quenching term {\mbox{$m_{\pi, vs}^2 = 1/2(m_{\pi}^2 + m_{\pi, ss}^2)$}} with $m_{\pi, ss}$ the pion mass corresponding to the sea quark mass in the global fit formula does not describe our lattice data well and the fit parameters $A_1$, $A_2$ do not have any signal in this case. With the partial-quenching
term included, one obtains $\left.G^\text{light-sea}_M(0)\right\vert_{\text{physical}} = - 0.147(33) \,\mu_N\,$. However, we include the second systematic error in our final result due to the possible inclusion of this partial quenching term in the global fit fit~(\ref{MFFglobal}). 

\begin{figure}[htbp]
\begin{center}
\includegraphics[height=6.0cm,width=9.0cm]{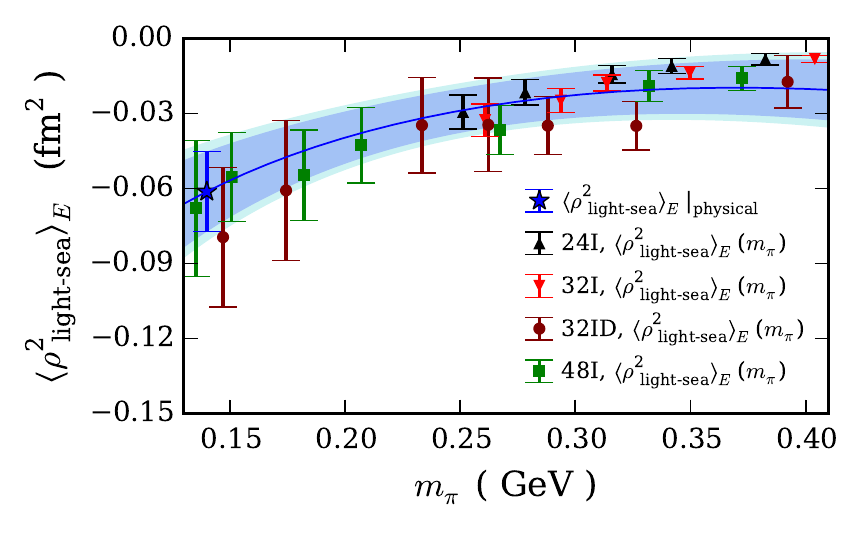}
\end{center}
\caption{\label{fig7}Light disconnected-sea quark charge radius at 24 quark masses on 24I, 32I, 48I, and 32ID ensembles as a function of the pion mass. The curved blue line in the figure shows the behavior in the infinite volume and continuum limit. The cyan band shows the combined statistical (blue band) and systematic uncertainties added in quadrature. The $\chi^2/\text{d.o.f.}$ of the fit is 0.46.}
\end{figure}

In Sec.~\ref{zexpEMFF}, we have obtained the light disconnected-sea quarks contribution to the charge radii using the $z$-expansion method by calculating the slope of $G^\text{light-sea}_E(Q^2)$ using the following definition:
\bea
\langle  \rho_\text{\,light-sea}^2\rangle_E \equiv -6 \left.\frac{dG^\text{light-sea}_E}{dQ^2}\right \vert_{Q^2=0}
\eea
where we used the symbol $\rho$ instead of the conventional symbol $r$ for the charge radius to emphasize the fact that charge factors are not yet included in the $G^\text{light-sea}_E(Q^2)$ form factor data. Using the charge radius values at 3 different volumes and lattice spacings and 24 valence-quark masses from four ensembles, we perform a simultaneous continuum and chiral extrapolation to obtain the final value of the charge radius using the following global fit formula:

\bea \la{radiusglobal}
\langle \rho_\text{\,light-sea}^2\rangle_E \, (m_{\pi}, m_{\pi,vs}, m_K, a, L)&=& A_0 + A_1\,\log\,(m_\pi) +A_2\, m_\pi^2 + A_3\,m_{\pi, vs}^2 \nn \\
&&  + A_4\,a^2 +A_5\, \sqrt{L}\,e^{-m_\pi L}.
\eea
 The chiral extrapolation in the empirical formula~(\ref{radiusglobal}) has been adopted from~\cite{Hemmert:1999mr} by replacing $m_K$ with $m_\pi$ and the volume correction similar to the pion charge radius correction has been obtained from~\cite{Tiburzi:2014yra}. In the continuum limit, we obtain
 \bea \la{radius}
\left.\langle \rho_\text{\,light-sea}^2\rangle_E \right\vert_{\text{physical}} = -0.061 (16)(11)(10)\, \text{fm}^2, 
\eea
and the fit parameters are: $A_1=0.077(24)$, $A_2=-0.280(99)$, $A_3=0.151(100)$, $A_4=-0.015(13)$, and $A_5=-0.054(58)$. The extraction of the charge radius from the FFs is sensitive to the lowest value of $Q^2$ and momentum transfer range of the data used, and also on the form of the fit. However, one wants to go to very low $Q^2$-values to extract the charge radii and the 48I ensemble has a lowest momentum transfer which is almost 4 times smaller than those of the 24I and 32I lattice data. It is seen from Fig.~\ref{fig7} that the uncertainties in the charge radii obtained from 48I and 32ID ensembles are large compared to the 24I and 32I ensemble results. We find that the uncertainty of the global fit result is almost equal to the uncertainty of charge radius obtained from the 48I ensemble at the valence quark mass equivalent to $m_\pi = 150$ MeV. 

It is important to note that the magnetic moment and charge radius results in Eqs.~(\ref{magmoment}) and (\ref{radius}) do not include charge factors. We define the magnetic moment in the unit of nucleon magneton $\mu_M$ and the charge radius as $\langle r^2\rangle_E$ with the proper charge factors included. After including the charge factors and using the results from~\cite{Sufian:2016pex} and Eqs.~(\ref{magmoment}) and (\ref{radius}) we obtain

\bea
\mu^\text{s}_M &=& -\frac{1}{3}G^s_M(0)\nn \\
&=&0.021(5) (3)\, \mu_N, \\
\mu^\text{light-sea}_M &=& (\frac{2}{3}-\frac{1}{3})G^\text{light-sea}_M(0)\nn \\
&=&- 0.043 (10) (08)\, \mu_N
\eea
Similarly, 
\bea
\langle \rho_\text{s}^2\rangle_E &=& -\frac{1}{3}\langle r_\text{s}^2\rangle_E\nn \\
&=& 0.0014 (05) (05)\, \text{fm}^2, \\
\langle \rho_\text{light-sea}^2\rangle_E &=& (\frac{2}{3}-\frac{1}{3}) \langle r_\text{light-sea}^2\rangle_E\nn \\
&=&-0.0203 (53) (49)\, \text{fm}^2.
\eea

Combining results with the strange quark magnetic moment and charge radius, we obtain the total contribution from the light and strange disconnected-sea quarks to the nucleon magnetic moment and charge radius

\bea \la{compare}
\mu_M\,\text{(DI)}&=& -0.022(11)(09)\, \mu_N, \\
\langle r^2\rangle_E\,\text{(DI)} &=& -0.019 (05) (05)\, \text{fm}^2.
\eea

Comparing with the PDG values of nucleon magnetic moments~\cite{PDG}, our results indicate that disconnected-sea quarks contribute~$\sim1\%$ to the nucleon magnetic moments, namely, a negative $0.8(5)\%$ and a $1.2(7)\%$ to the proton and neutron magnetic moments, respectively. Keeping in mind that there is a $4\%$ discrepancy between the measurement of proton charge radius from the muonic Lamb shift experiment and the electron-proton scattering experiments, our finding in the present work reveals that the lattice calculation of the DI gives a negative 2.5(9)\% contribution to the proton mean square charge radius. This is about 1/3 of the discrepancy between the proton mean square charge radii measured in the electron-proton scattering and the muonic atom. Thus, it is important to have the DI included when the lattice calculation of the proton charge radius is carried out. Although a complete lattice QCD calculation including the connected and disconnected insertions at the physical point is required to draw any definitive conclusion about the accurate percentage of the disconnected-sea quarks contribution to a proton charge radius, this calculation clearly indicates that there will be a shift toward a smaller value of the proton charge radius when the light disconnected-sea quarks contribution is included. However, the disconnected-sea quarks contribution to the neutron mean square charge radius can have a significant effect, namely 16.3(6.1)\% compared to the experimental neutron mean square charge radius $\langle r_n^2\rangle = -0.1161(22)$ fm$^2$~\cite{PDG}, in obtaining a value closer to the experimental value. 

\begin{figure}[htbp]
  \centering
  \subfigure[]{\includegraphics[width=0.48\textwidth]{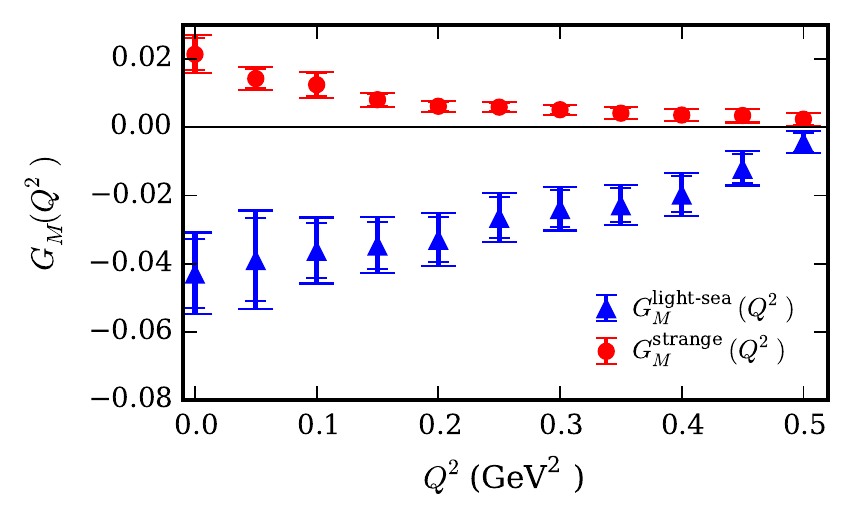}\la{8a}}
  \subfigure[]{\includegraphics[width=0.48\textwidth]{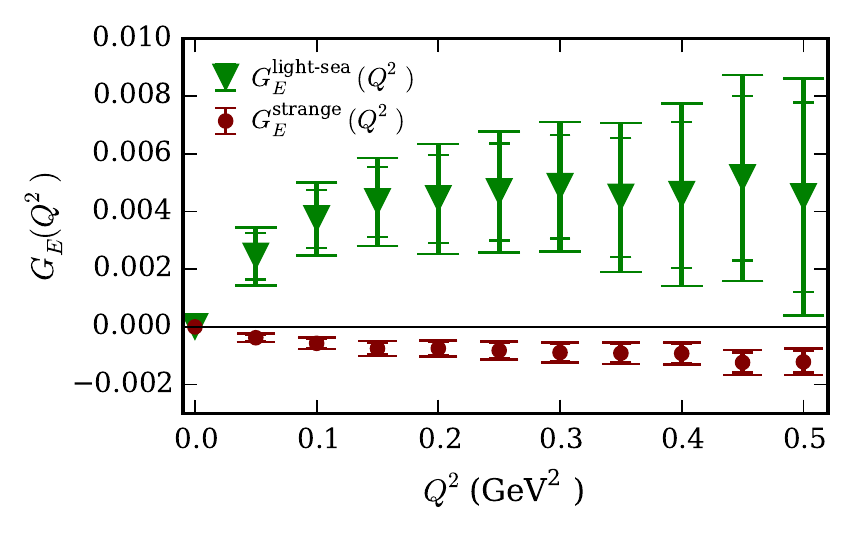}\la{8b}}
  \caption{\label{fig8}Light and strange disconnected-sea quarks contributions to the nucleon electromagnetic form factors at the physical point and in the continuum limit. Charge factors are included in the form factor calculations. The outer error bars in the data points include the systematic uncertainties in the calculations.}
\end{figure}

From the $z$-expansion fit parameters in Sec.~\ref{zexpEMFF}, we can now interpolate the light and strange disconnected-sea quarks contributions to the nucleon electromagnetic form factors. Although the largest available momentum transfer we have on the 24I and 32I ensemble is $Q^2\sim 1.3$ GeV$^2$, the largest momentum transfer available on the 48I ensemble is $Q^2~\sim 0.5$ GeV$^2$. Therefore, we note that the extrapolation of the nucleon EMFF starts to break down after $Q^2~\sim 0.4$ GeV$^2$ for the 48I ensemble and we constrain the extrapolations of the 48I ensemble EMFF up to $Q^2=0.5$ GeV$^2$. The global fit results of the strange quark EMFFs have been obtained from~\cite{Sufian:2016vso} and we use similar empirical formulas as Eqs.~(\ref{MFFglobal}) and (\ref{radiusglobal}) to estimate the light-sea quarks contribution to the nucleon EMFF in the continuum limit and at the physical point. The contributions of $G_{E,M}(Q^2)$\,\text{(DI)} to the nucleon electromagnetic form factors appear with charge factors. Therefore, we present the results in Fig.~\ref{fig8} with systematics included and also include charge factors in the form factor calculations so that the sign and magnitude of the disconnected-quarks contributions to the nucleon EMFFs can directly be compared to the nucleon total EMFFs. These results will be combined with the connected insertion calculation of the nucleon EMFFs in our future work to obtain a complete description of the nucleon EMFF from first-principles calculation. 

\section{Conclusion} \la{conclude}
In this calculation, we have uncovered the practical importance of including the disconnected quark loops contribution to the nucleon magnetic moment and charge radius. In particular, in accord with the analysis, we find that the light and strange disconnected sea-quarks contribution to the nucleon charge radius can have an important impact to reconcile lattice QCD estimates with experimental measurements. A negative 2.5(9)\% contribution to the proton mean square charge radius from the disconnected-sea quarks should have an impact on the ``proton charge radius puzzle'' where the discrepancy in the mean square charge radius of $\sim8\%$ is of the same order. It is seen for the first time that the disconnected quarks can shift the neutron mean square charge radius calculation toward the experimental value by about 16\%. Especially, because the neutron electric form factor calculation on the lattice is noisy and the connected-insertion-only quark contribution is smaller than the experimental $Q^2$ behavior, the disconnected quark loops cannot be ignored for an accurate estimation of the neutron form factors at low $Q^2$ on the lattice. Our main focus of this calculation was to show that 1) the disconnected-sea quarks contribution to the nucleon properties at low $Q^2$ is of significant importance and 2) numerical simulation with controlled systematics and at the physical pion mass can generate a better theoretical understanding of various nucleon properties.

\begin{acknowledgments}

We thank the RBC and UKQCD Collaborations for providing their DWF gauge configurations. This work is supported in part by the U.S. DOE Grant $\text{No.}$ $\text{DE-SC}0013065$. This research used resources of the Oak Ridge Leadership Computing Facility at the Oak Ridge National Laboratory, which is supported by the Office of Science of the U.S. Department of Energy under Contract No. DE-AC05-00OR22725. This work used Stampede time under the Extreme Science and Engineering Discovery Environment (XSEDE)~\cite{XSEDE}, which is supported by National Science Foundation Grant No. ACI-1053575. We also thank National Energy Research Scientific Computing Center (NERSC) for providing HPC resources that have contributed to the research results reported within this paper. We acknowledge the facilities of the USQCD Collaboration used for this research in part, which are funded by the Office of Science of the U.S. Department of Energy.
\end{acknowledgments}

\providecommand{\href}[2]{#2}
\begingroup\raggedright

\endgroup

\end{document}